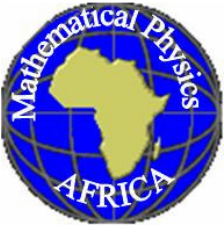

# Two – Dimensional Quantum (4,4) Null Superstring in de Sitter Space

F. Assaoui* and T. Lhallabi**

*The Abdus Salam International Centre for Theoretical Physics, Miramare-Trieste, Italy*

*Abstract*

The (4,4) null superstring equations of motions and constraints on de Sitter space are given by using the harmonic superspace. These are solved explicitly by performing a perturbative expansion of the (4,4) superstring coordinates in powers of $c^2$, the world-sheet speed of light. The analytic expressions of the zeroth and first order solutions are determined. On the other hand, we study the quantization of the (4,4) null superstring in de Sitter space and we describe its superalgebra.

## *I - Introduction*

A consistent quantum theory of gravity is the strongest motivation for string theory and hence to study strings in curved spacetime [1]. A sensible theory of quantum gravity is necessarily a part of the unified theory of all interactions in fact that [2] a quantum theory of gravity must be a theory able to describe all physics below the Planck scale $M_{Plank} = \frac{\eta c}{G} = 1.22\,10^{16}\,Tev$ [1]. As a first step on the understanding of quantum gravitational phenomena in a string framework, H.J. de Vega and N. Sanchez have started in 1987 a programme of string quantization on curved spacetime [2,3]. The investigation of strings on curved spacetime is currently the best framework to study the physics of gravitation [4].

The classical and the quantum string propagation in curved spacetime is a very important subject. The investigations in this domain are relevant for the physics of quantum gravitation as well as for the understanding of the cosmic string models in cosmology [2,3,5]. It is well known that the string equations of motion in curved spacetime constitute a complicated system of non linear coupled second order partial differential equations which in general, are non integrable [6]. In that case of tensionless string [7], this situation is simplified since the null strings, similarly to the massless point particles, essentially sweep out the light cone, and their equations of motion are essentially just geodesic equations of general relativity appended by an additional constraint. General relativistic first integrals for point particles are known for most of the symmetric spacetimes and we can apply them to null strings with almost no hesitation. Then, depending on the assumed shape of null string, in principle, one can solve the null string equations of motion in many cases. A lot of such explicit solutions have already been found, but they were restricted to just the most symmetric spacetimes such as Minkowski, de Sitter, anti-de Sitter, Schwarzschild and Robertson-Walker ones [2,3,8,9]. Recently, the string theory in Anti de Sitter space has using the light cone gauge [10]. They have computed the free spectrum and they show that at zero order in the 't Hooft coupling the corrections to the conformal dimension are finite.

The frequencies of the string modes are proportional to $T_0$ and the length of the string scales with $1/\sqrt{T_0}$ such that $\sqrt{T_0}$ is the energy scale of string with tension $T_0$. Therefore, the gravitational field provides another length scale, the curvature radius of the spacetime $R_c$. For a string moving in a gravitational field a useful parameter is the dimensionless constant $g = R_c \sqrt{T_0}$. Large values of $g$ imply weak gravitational field. We may reach large values of $g$ by letting $T_0 \to \infty$. In this limit the string shrinks to a point and a suitable expansion has been proposed [1,2,4,8]. In the opposite limit, small values of $g$, we encounter strong gravitational fields and it is appropriate to consider $T_0 \to 0$. Furthermore, a systematic expansion in terms of string tension has been presented [11] and at zeroth order the null string has been obtained [7,12] where every point of the string moves independently along a null geodesic. Purpose of this work is the discussion of the (4,4) null superstring evolution in de Sitter space by using the techniques of the curved harmonic superspace [13].

The outline of this paper is as follows: In section II, we start by presenting the (4,4) null superstring equations of motion in the curved harmonic superspace. The corresponding super constraints are then obtained by using the expansion of the analytic superfield. In section III, we study the quantization of the (4,4) null superstring in de Sitter space. The constraint relating the initial shape and momentum conjugates are solved and the corresponding superalgebra is obtained. Finally, in section IV, we make concluding remarks and discuss our results.

## II - Null Superstring in Curved (4,4) Superspace

In order to study supergravity in the context of superstring theory, let us consider superstring propagating in curved two dimensional harmonic superspace with the following action

$$S = -\frac{T}{2} \int d^2\sigma \, d^2\theta_+^+ \, d^2\theta_-^- \, du \, \sqrt{-h} \, h^{--++}(z_A, u) \, G_{ab}(\Omega) \, D^{++}\Omega^a \, D^{--}\Omega^b \qquad (2.1)$$

where $0 \leq a,b \leq d-1$, $T = (2\Pi\alpha')^{-1}$ is the string tension while $h^{--++}(z_A, u)$ describes the geometry of the two dimensional world-sheet [14] and $G_{ab}(\Omega)$ describes that of the target space. Furthermore, we have to take the conformal gauge namely

$$h^{--++}(z_A,u) = e^{\Phi(z_A,U)} \eta^{--++} \tag{2.2}$$

which allows to discuss the limit $T \to 0$ [12], where $\Phi(z_A,u)$ is an arbitrary function and

$$\eta^{--++} = diag(-1,+1)$$

This leads to the classical equations of motion

$$D^{++}D^{--}\Omega^a + \Gamma^a_{bc}(\Omega) D^{++}\Omega^b D^{--}\Omega^c = 0 \tag{2.4}$$

with

$$\Gamma^a_{bc}(\Omega) = \frac{1}{2} G^{ab}\{G_{ec,b} + G_{be,c} - G_{bc,e}\} \tag{2.5}$$

is the Christofell symbols associate to the metric $G_{ab}(\Omega)$ where $G_{ab,c}(\Omega)$ indicates differentiation with respect to $\Omega^c$. By using the expansion of the analytic superfield $\Omega^a$ namely:

$$\Omega^a = \omega^a + \theta^+_+\psi^{-a}_- + \theta^-_-\chi^{+a}_+ + (\theta^+_+)^2 b^{--a}_{--} + (\theta^-_-)^2 C^{++a}_{++} + \theta^+_+\theta^-_-\rho^a + (\theta^+_+)^2\theta^-_-\eta^{-a}_- + \theta^+_+(\theta^-_-)^2 \xi^{+a}_+$$
$$+ (\theta^+_+)^2(\theta^-_-)^2 m^a$$
$$\tag{2.6}$$

the equation (2.4) leads to the following constraints

$$\partial^{++}\partial^{--}\omega^l + \Gamma^l_{ak}(\omega)\partial^{++}\omega^a\partial^{--}\omega^k = 0 \tag{2.7.1}$$

$$\partial^{++}\partial^{--}\psi^{-l}_- + \Gamma^l_{ak}(\omega)[\partial^{++}\omega^a\partial^{--}\psi^{-k}_- + \partial^{++}\psi^{-a}_-\partial^{--}\omega^k] = 0 \tag{2.7.2}$$

$$\partial^{++}\partial^{--}\chi^{+l}_+ + \Gamma^l_{ak}(\omega)[\partial^{++}\omega^a\partial^{--}\chi^{+k}_+ + \partial^{++}\chi^{+a}_+\partial^{--}\omega^k] = 0 \tag{2.7.3}$$

$$\partial^{++}\partial^{--}\rho^l + \Gamma^l_{ak}(\omega)[\partial^{++}\omega^a\partial^{--}\rho^k + \partial^{++}\psi^{-a}_-\partial^{--}\chi^{+k}_+ + \partial^{++}\chi^{+k}_+\partial^{--}\psi^{-a}_- + \partial^{++}\rho^a\partial^{--}\omega^k] = 0 \tag{2.7.4}$$

$$\partial^{++}\partial^{--}b^{--l}_{--} - 2\partial_{--}\partial^{--}\omega^l + \Gamma^l_{ak}(\omega)[\partial^{++}\omega^a\partial^{--}b^{--k}_{--} + \partial^{++}b^{--a}_{--}\partial^{--}\omega^k - 2\partial_{--}\omega^l\partial^{--}\omega^k$$
$$+ \frac{1}{2}\partial^{++}\psi^{-a}_-\partial^{--}\psi^{-k}_-] = 0 \tag{2.7.5}$$

$$\partial^{++}\partial^{--}C^{++l}_{++} - 2\partial^{++}\partial_{++}\omega^l + \Gamma^l_{ak}(\omega)[\partial^{++}\omega^a\partial^{--}C^{++k}_{++} - 2\partial^{++}\omega^a\partial_{++}\omega^k + 2\partial^{++}\chi^{+a}_+\partial^{--}\chi^{-k}_-$$
$$+ \partial^{++}C^{++a}_{++}\partial^{--}\omega^k] = 0 \tag{2.7.6}$$

$$\partial^{++}\partial^{--}\eta^{-l}_- - 2\partial_{--}\partial^{--}\chi^{+l}_+ + \Gamma^l_{ak}(\omega)[\partial^{++}\omega^a\partial^{--}\eta^{-k}_- + \frac{1}{2}\partial^{++}\psi^{-a}_-\partial^{--}\rho^k + \partial^{++}\chi^{+a}_+\partial^{--}b^{--k}_{--}$$
$$+ \partial^{--}b^{--a}_{--}\partial^{++}\chi^{+k}_+ + \partial^{++}\rho^a\partial^{--}\psi^{-k}_- + \partial^{++}\eta^{-a}_-\partial^{--}\omega^k \tag{2.7.7}$$
$$- 2\partial_{--}\chi^{+a}_+\partial^{--}\omega^k - 2\partial_{--}\omega^a\partial^{--}\chi^{+l}_+] = 0$$

$$\partial^{++}\partial^{--}\xi^{+l}_+ - 2\partial^{++}\partial_{++}\psi^{-l}_- + \Gamma^l_{ak}(\omega)[\partial^{++}\omega^a\partial^{--}\xi^{+k}_+ - \partial^{++}\omega^a\partial_{++}\psi^{-k}_- + \partial^{++}\psi^{-a}_-\partial^{--}C^{++k}_{++}$$
$$- 2\partial^{++}\psi^{-a}_-\partial_{++}\omega^k + \partial^{++}\chi^{+a}_+\partial^{--}\rho^k + \partial^{++}C^{++a}_{++}\partial^{--}\psi^{-k}_- \tag{2.7.8}$$
$$+ \partial^{++}\rho^a\partial^{--}\chi^{+k}_+ + \partial^{++}\xi^{+a}_+\partial^{--}\omega^k] = 0$$

$$\partial^{++}\partial_{++}b_{--}^{-\:-l} + \partial_{--}\partial^{--}C_{++}^{++l} - 2\partial_{--}\partial_{++}\omega^l - \frac{1}{2}\partial^{++}\partial^{--}m^l - \frac{1}{2}\Gamma^l_{ak}(\omega)[\partial^{++}\psi_-^{-a}\partial^{--}\xi_+^{+l} - 2\partial^{++}\omega^a\partial_{++}b_{--}^{-\:-k}$$

$$-2\partial^{++}\psi_+^{+a}\partial_{++}\psi_-^{-k} + \partial^{++}\chi_+^{+a}\partial^{--}\eta_-^{-k}$$

$$+\partial^{++}b_{--}^{-\:-a}\partial^{--}C_{++}^{++k} - 2\partial^{++}b_{--}^{-\:-a}\partial_{++}\omega^k$$

$$+\partial^{++}C_{++}^{++a}\partial^{--}b_{--}^{-\:-k} + \partial^{++}\rho^a\partial^{--}\rho^k + \partial^{--}\eta_-^{-a}\partial^{++}\chi_+^{+k}$$

$$+\partial^{++}\xi_+^{+a}\partial^{--}\psi_-^{-k} + \partial^{++}m^a\partial^{--}\omega^k$$

$$-2\partial_{--}\omega^a\partial^{--}C_{++}^{++k} + 4\partial_{--}\omega^a\partial_{++}\omega^a$$

$$-2\partial_{--}\chi_+^{+a}\partial^{--}\chi_+^{+k} - 2\partial_{--}C_{++}^{++a}\partial^{--}\omega^k] = 0$$

(2.7.9)

Supplemented by the constraints

$$\partial^{--}\partial^{++}\omega^l = \partial^{++}\partial^{--}\omega^l \qquad (1)$$
$$\partial^{--}\partial^{++}\psi_-^{-l} = \partial^{++}\partial^{--}\psi_-^{-l} \qquad (2)$$
$$\partial^{--}\partial^{++}\rho^l = \partial^{++}\partial^{--}\rho^l \qquad (3)$$
$$\partial^{--}\partial^{++}b_{--}^{-\:-l} = \partial^{++}\partial^{--}b_{--}^{-\:-l} \qquad (4)$$
$$\partial^{--}\partial^{++}C_{++}^{++l} = \partial^{++}\partial^{--}C_{++}^{++l} \qquad (5) \qquad (2.8)$$
$$\partial^{--}\partial^{++}\eta_-^{-l} = \partial^{++}\partial^{--}\eta_-^{-l} \qquad (6)$$
$$\partial^{--}\partial^{++}\xi_+^{+l} = \partial^{++}\partial^{--}\xi_+^{+l} \qquad (7)$$

$$\partial^{--}\partial_{--}C_{++}^{++l} + \partial_{++}\partial^{++}b_{--}^{-\:-l} - 2\partial_{++}\partial_{--}\omega^l - \frac{1}{2}\partial^{--}\partial^{++}m^l = -\frac{1}{2}\partial^{++}\partial^{--}m^l + \partial^{++}\partial_{++}b_{--}^{-\:-l}$$
$$+ \partial_{--}\partial^{--}C_{++}^{++l} - 2\partial_{--}\partial_{++}\omega^l \qquad (8)$$

which is deduced from the condition

$$D^{++}D^{--}\Omega^l = D^{--}D^{++}\Omega^l \qquad (2.9)$$

since $D^0\Omega^l = 0$ [15]. Furthermore, the consistency condition

$$(D^{++})^2\Omega^a = 0 \qquad (2.10)$$

leads to

$$(\partial^{++})^2\omega^a = 0 \qquad (1)$$
$$(\partial^{++})^2\psi_-^{-a} = 0 \qquad (2)$$
$$(\partial^{++})^2\chi_+^{+a} = 0 \qquad (3) \qquad (2.11)$$
$$\partial^{++}b_{--}^{-\:-a} = 4\partial_{--}\omega^a \qquad (4)$$
$$\partial^{++}\eta_-^{-a} = 4\partial_{--}\chi_-^{+a} \qquad (5)$$
$$\partial^{++}m^a = 4\partial_{--}C_{++}^{++a} \qquad (6)$$

Consequently, the field components $\omega^a, \psi_-^{-a}$ and $\chi_+^{+a}$ are given by the following expressions

$$\omega_a = \omega_a^0 + \omega_a^{\alpha\beta} u_\alpha^+ u_\beta^-$$
$$\psi_{-a}^- = \psi_{-a}^\alpha u_\alpha^-$$
$$\chi_{+a}^+ = \chi_{+a}^\alpha u_\alpha^+ + \chi_{+a}^{\alpha\beta\gamma} u_\alpha^+ u_\beta^+ u_\gamma^-$$

(2.12)

Let us note that the equations (2.7.1) until (2.7.4) do not contain the spatial derivatives and they do not contribute to the null superstring method which consist of inserting the expansion

$$\Omega^a(Z_A,u) = V^a(Z_A,u) + c^2 U^a(Z_A,u) + O(c^4) \tag{2.13}$$

Whereas, in the equations (2.7.5) until (2.7.9) the spatial derivative is appearing and the use of the constraints (2.8) and (2.11) allows rewriting them respectively as follows

$$2\partial_{--}(\partial^{--}\omega^l) + \Gamma_{ak}^l(\omega)[\partial^{++}\omega^a\partial^{--}b_{--}^{-k} + 2\partial_{--}\omega^a\partial^{--}\omega^k + \frac{1}{2}(\partial^{--}\psi_-^{-a})(\partial^{--}\psi_-^{-k})] = 0$$

(2.14.1)

$$\partial^{++}\partial^{--}C_{++}^{++l} - 2\partial^{++}\partial_{++}\omega^l + \Gamma_{ak}^l(\omega)[\partial^{++}\omega^a\partial^{--}C_{++}^{++k} - 2\partial^{++}\omega^a\partial_{--}\omega^k$$
$$+ \partial^{++}\chi_+^{+a}\partial^{--}\chi_+^{+k} + \partial^{++}C_{--}^{--a}\partial^{--}\omega^k] = 0$$

(2.14.2)

$$2\partial^{--}\partial_{--}\chi_-^{+l} + \Gamma_{ak}^l(\omega)[\partial^{++}\omega^a\partial^{--}\eta_-^{-k} + \frac{1}{2}\partial^{++}\psi_-^{-a}\partial^{--}\rho^k + \frac{1}{2}\partial^{++}\rho^a\partial^{--}\psi_-^{-k} + \partial^{++}\chi_+^{+a}\partial^{--}b_{--}^{-k}$$
$$+ 2\partial_{--}\omega^a\partial^{--}\chi_+^{+k} + 2\partial_{--}\chi_-^{+a}\partial^{--}\omega^k] = 0$$

(2.14.3)

$$2\partial^{++}\partial_{++}\psi_-^{-l} - \partial^{++}\partial^{--}\xi_+^{+l} - \Gamma_{ak}^l(\omega)[\partial^{++}\omega^a\partial^{--}\xi_+^{+k} - 2\partial^{++}\omega^a\partial_{++}\psi_-^{-k} + \partial^{++}\psi_-^{-a}\partial^{--}C_{++}^{++k}$$
$$- 2\partial^{++}\psi_-^{-a}\partial_{++}\omega^k + \partial^{++}\chi_+^{+a}\partial^{--}\rho^k + \partial^{++}C_{++}^{++a}\partial^{--}\psi_-^{-k}$$
$$+ \partial^{++}\rho^a\partial^{--}\chi_+^{+k} + \partial^{++}\xi_+^{+a}\partial^{--}\omega^k] = 0$$

(2.14.4)

$$4\partial_{--}\partial_{++}\omega^l - 2\partial^{--}\partial_{--}C_{++}^{++l} - \Gamma_{ak}^l(\omega)[-2\partial^{++}\omega^a\partial_{++}b_{--}^{-k} + \partial^{++}\psi_-^{-a}\partial^{--}\xi_+^{+l} + \partial^{++}\xi_+^{+a}\partial^{--}\psi_-^{-k}$$
$$- 2\partial^{++}\psi_-^{-a}\partial_{++}\psi_-^{-k} + \partial^{++}\chi_+^{+a}\partial^{--}\eta_-^{-k} + 2\partial_{--}\omega^a\partial^{--}C_{++}^{++k}$$
$$- 2\partial_{--}C_{++}^{++a}\partial^{--}\omega^k - 4\partial_{--}\omega^a\partial_{++}\omega^k + \partial^{++}C_{++}^{++a}\partial^{--}b_{--}^{-k}$$
$$+ \partial^{++}\rho^a\partial^{--}\rho^k + 2\partial_{--}\chi_+^{+a}\partial^{--}\chi_+^{+k} = 0$$

(2.14.5)

On the other hand, the examination of the following consistency constraint [15]

$$(D^{--})^2 \Omega^a = 0 \tag{2.15}$$

leads to

$$(\partial^{--})^2 \omega^l = 0 = (\partial^{--})^2 \rho^l \quad (1)$$
$$(\partial^{++})^2 \psi_-^{-l} = 0 = (\partial^{++})^2 \chi_+^{+l} \quad (2)$$
$$(\partial^{++})^2 b_{--}^{-l} = 0 = (\partial^{--})^2 \eta_-^{-l} \quad (3)$$
$$\partial^{--} C_{++}^{++l} = 4\partial_{++}\omega^l \quad (4) \quad (2.16)$$
$$\partial^{++} \xi_+^{+a} = 4\partial_{++}\psi_-^{-l} \quad (5)$$
$$\partial^{--} m^l = 4\partial_{++}b_{--}^{-l} \quad (6)$$

The constraints allow concluding that

$$\psi_{-}^{-a} = \psi_{-}^{i,a} u_{i}^{-} \qquad (1)$$
$$\chi_{+}^{+a} = \chi_{+}^{i,a} u_{i}^{+} \qquad (2)$$
$$\partial^{++}\psi_{-}^{-a} = \psi_{-}^{+a} \qquad (3)$$
$$\partial^{--}\chi_{+}^{+a} = \chi_{+}^{-a} \qquad (4)$$

(2.17)

Therefore

$$\partial^{--}\psi_{-}^{-a} = 0$$
$$\partial^{++}\chi_{+}^{+a} = 0$$

(2.18)

By using equations (2.16), (2.17) and (2.18) the constraints (2.7) take the following forms

$$\psi_{-}^{-l} + \Gamma_{ak}^{l}(\omega)\,\partial^{--}\omega^{k}\,\psi_{-}^{+a} = 0$$

$$\chi_{+}^{+l} + \Gamma_{ak}^{l}(\omega)\,\partial^{++}\omega^{a}\,\chi_{-}^{-k} = 0$$

$$\partial^{++}\partial^{--}\rho^{l} + \Gamma_{ak}^{l}(\omega)[\partial^{++}\omega^{a}\partial^{--}\rho^{k} + \psi_{-}^{+a}\chi_{+}^{-k} + \partial^{++}\rho^{a}\partial^{--}\omega^{k}] = 0$$

$$2\partial_{--}\partial^{--}\omega^{l} + \Gamma_{ak}^{l}(\omega)[\partial^{++}\omega^{a}\partial^{--}b_{--}^{-k} + 2\partial_{--}\omega^{a}\partial^{--}\omega^{k}] = 0$$

$$2\partial^{++}\partial_{++}\omega^{l} + \Gamma_{ak}^{l}(\omega)[2\partial^{++}\omega^{a}\partial_{++}\omega^{k} + \partial^{++}C_{++}^{++a}\partial^{--}\omega^{k}] = 0 \qquad (2.19)$$

$$2\partial_{--}\chi_{+}^{-l} + \Gamma_{ak}^{l}(\omega)[\partial^{++}\omega^{a}\partial^{--}\eta_{-}^{-k} + \psi_{-}^{+a}\partial^{--}\rho^{k} + 2(\partial_{--}\omega^{a})\chi_{+}^{-k} + 2\partial_{--}\chi_{+}^{+a}\partial^{--}\omega^{k}] = 0$$

$$2\partial_{++}\psi_{-}^{+l} + \Gamma_{ak}^{l}(\omega)[2\partial^{++}\omega^{a}\partial_{++}\psi_{-}^{-k} + 2\partial\psi_{-}^{+a}\partial_{++}\omega^{k} + \partial^{++}\rho^{a}\partial^{--}\chi_{+}^{-k} + \partial^{++}\xi_{+}^{+a}\partial^{--}\omega^{k}] = 0$$

$$\partial_{--}\partial_{++}\omega^{l} + \Gamma_{ak}^{l}(\omega)[\partial_{--}\omega^{a}\partial_{++}\omega^{k} + \frac{1}{2}\psi_{-}^{+a}\partial_{++}\psi_{-}^{-k} + \frac{1}{2}\chi_{+}^{-k}\partial_{--}\chi_{+}^{+a} + \frac{1}{4}\partial^{++}\rho^{a}\partial^{--}\rho^{k} - \frac{1}{2}\partial^{++}\omega^{a}\partial_{++}b_{--}^{-k}$$

$$+ \frac{1}{2}\partial_{--}C_{++}^{++a}\partial^{--}\omega^{k} + \frac{1}{4}\partial^{++}C_{++}^{++a}\partial^{--}b_{--}^{-k} = 0$$

The introduction of the light cone variables $y^{\pm\pm}$ on the world-sheet namely

$$y^{\pm\pm} = \tau \pm \sigma$$
$$\partial_{\pm\pm} = \frac{\partial}{\partial y^{\mu\mu}} \equiv \partial_{\tau} \mu \, \partial_{\sigma}$$

(2.20)

with

$$\sigma = \frac{1}{c}\sigma$$
$$\partial_{\pm\pm} = \partial_{\tau} \mu \, c\partial_{\sigma}$$

transforms the equations (2.19) to the following

$$\partial^{--}\dot{\omega}^{l} + c\partial^{--}\omega'^{l} + \frac{1}{2}\Gamma_{ak}^{l}(\omega)[\partial^{++}\omega^{a}\partial^{--}b_{--}^{-k} + 2\dot{\omega}^{a}\partial^{--}\omega^{k} + 2c\omega'^{a}\,\partial^{--}\omega'^{k}] = 0 \qquad (2.21.1)$$

$$\partial^{--}\dot{\omega}^{l} - c\partial^{++}\omega'^{l} + \frac{1}{2}\Gamma_{ak}^{l}(\omega)[2\dot{\omega}^{k}\partial^{++}\omega^{a} + \partial^{++}C_{++}^{++a}\partial^{--}\omega^{k} - 2c\omega'^{a}\,\partial^{++}\omega'^{k}] = 0 \qquad (2.21.2)$$

$$\chi_{-}^{\prime-l} + c\chi_{+}^{\prime-l} + \frac{1}{2}\Gamma_{ak}^{l}(\omega)[2\partial\omega^{a}\chi_{+}^{-k} + 2C\omega^{\prime a}\chi_{+}^{-k} + 2c\chi_{+}^{+a}\partial^{--}\omega^{k} + 2c\chi_{+}^{\prime+a}\partial^{--}\omega^{k}$$
$$+\partial^{++}\omega^{a}\partial^{--}\eta_{-}^{-k} + \psi_{-}^{+a}\partial^{--}\rho^{k}] = 0 \tag{2.21.3}$$

$$\psi_{-}^{\prime+l} - c\psi_{-}^{\prime+l} + \frac{1}{2}\Gamma_{ak}^{l}(\omega)[2(\partial^{++}\omega)\psi_{-}^{-k} - 2c(\partial^{++}\omega^{a})\psi_{-}^{\prime-k} - 2\psi_{-}^{+a}\partial\omega^{k} - 2c\psi_{-}^{+a}\omega^{\prime k}$$
$$+ (\partial^{++}\rho^{a})\chi_{+}^{-k} + \partial^{++}\xi_{+}^{+a}(\partial^{--}\omega^{k})] = 0 \tag{2.21.4}$$

$$\partial\omega^{l} - c^{2}\omega^{\prime\prime l} + \frac{1}{2}\Gamma_{ak}^{l}(\omega)[4\partial\omega^{a}\partial\omega^{k} - 4c^{2}\omega^{\prime l}\omega^{\prime k} + 2\psi_{-}^{+a}\psi_{-}^{-k} - 2c\psi_{-}^{+a}\psi_{-}^{\prime-k} - 2\chi_{+}^{-k}\partial\chi_{+}^{+a}$$
$$+ 2c\chi_{+}^{-k}\chi_{+}^{\prime+a} - 2\partial^{++}\omega^{a}(b_{--}^{-k} - cb_{--}^{\prime-k}) + 2(C_{++}^{++a} - cC_{++}^{\prime++a})\partial^{--}\omega^{a}$$
$$+ \partial^{++}\rho^{a}\partial^{--}\rho^{k} + \partial^{++}C_{++}^{++a}\partial^{--}b_{--}^{-k}] = 0 \tag{2.21.5}$$

Furthermore, the variation of the action (2.1) with respect to the two-dimensional metric $h^{--++}(z_A,u)$ gives

$$G_{ab}(\Omega)D^{++}\Omega^{a}D^{--}\Omega^{b} = 0 \tag{2.22}$$

which yields the constraints

$$G_{ab}(\omega)\partial^{++}\omega^{a}\partial^{--}\omega^{b} = 0 \tag{2.23.1}$$
$$G_{ab}(\omega)[\partial^{++}\omega^{a}\partial^{--}\psi_{-}^{-a} + \partial^{++}\psi_{-}^{-a}\partial^{--}\omega^{b}] = 0 \tag{2.23.2}$$
$$G_{ab}(\omega)[\partial^{++}\omega^{a}\partial^{--}\chi_{+}^{+a} + \partial^{++}\chi_{+}^{+a}\partial^{--}\omega^{b}] = 0 \tag{2.23.3}$$
$$G_{ab}(\omega)[\partial^{++}\omega^{a}\partial^{--}b_{--}^{-b} + \partial^{++}\psi_{-}^{-a}\partial^{--}\psi_{-}^{-b} + \partial^{++}b_{--}^{-b}\partial^{--}\omega^{b} - 2\partial_{--}\omega^{a}\partial^{--}\omega^{b}] = 0 \tag{2.23.4}$$
$$G_{ab}(\omega)[\partial^{++}\omega^{a}\partial^{--}C_{++}^{++b} - 2\partial^{++}\omega^{a}\partial_{--}\omega^{b} + \partial^{++}\chi_{+}^{+a}\partial^{--}\chi_{+}^{+b} + \partial^{++}C_{++}^{++a}\partial^{--}\omega^{b}] = 0 \tag{2.23.5}$$
$$G_{ab}(\omega)[\partial^{++}\omega^{a}\partial^{--}\rho^{b} + \partial^{++}\psi_{-}^{-a}\partial^{--}\chi_{+}^{+b} + \partial^{++}\chi_{+}^{+a}\partial^{--}\psi_{-}^{-b} + \partial^{++}\rho^{a}\partial^{--}\omega^{b}] = 0 \tag{2.23.6}$$
$$G_{ab}(\omega)[\partial^{++}\omega^{a}\partial^{--}\eta_{-}^{-b} + \partial^{++}\psi_{-}^{-a}\partial^{--}\rho^{b} + \partial^{++}\chi_{+}^{+a}\partial^{--}b_{--}^{-b} + \partial^{++}b_{--}^{-b}\partial^{--}\chi_{+}^{+b}$$
$$+ \partial^{++}\rho^{a}\partial^{--}\psi_{-}^{-b} + \partial^{++}\eta_{-}^{-a}\partial^{--}\omega^{b} - 2\partial_{--}\chi_{+}^{+a}\partial^{--}\omega^{b} - 2\partial_{--}\omega^{a}\partial^{--}\chi_{+}^{+b}] = 0 \tag{2.23.7}$$

$$G_{ab}(\omega)[\partial^{++}\omega^{a}\partial^{--}m^{b} - 2\partial^{++}\omega^{a}\partial_{++}b_{--}^{-b} + \partial^{++}\psi_{-}^{-a}\partial^{--}\xi_{+}^{+b} - 2\partial^{++}\psi_{-}^{-a}\partial_{--}\psi_{-}^{-b} + \partial^{++}\chi_{+}^{+a}\partial^{--}\eta_{-}^{-a}$$
$$+ \partial_{++}b_{--}^{-a}\partial^{--}C_{++}^{++b} - 2\partial^{++}b_{--}^{-a}\partial_{++}\omega^{b} + \partial^{++}C_{++}^{++a}\partial^{--}b_{--}^{-b} + \partial^{++}\rho^{a}\partial^{--}\rho^{b} + \partial^{++}\eta_{-}^{-a}\partial^{--}\chi_{+}^{+b}$$
$$+ \partial^{++}\xi_{+}^{+a}\partial^{--}\psi_{-}^{-a} + \partial^{++}m^{a}\partial^{--}\omega^{b} - 2\partial_{--}\omega^{a}\partial^{--}C_{++}^{++b} + 4\partial_{--}\omega^{a}\partial_{++}\omega^{b}$$
$$- 2\partial_{--}\chi_{+}^{+a}\partial^{--}\chi_{+}^{+b} - 2\partial_{--}C_{++}^{++a}\partial^{--}\omega^{b}] = 0 \tag{2.23.8}$$

In order to simplify this study let us take just the first terms of the field expansions in harmonic variables namely

$$\omega^a(\sigma,\tau,u) = \omega^a(\sigma,\tau) + f_1^a(\sigma,\tau,u)$$
$$\psi^-_{-a}(\sigma,\tau,u) = \psi^\alpha_{-a}(\sigma,\tau)u^-_\alpha$$
$$\chi^+_{+a}(\sigma,\tau,u) = \chi^\alpha_{+a}(\sigma,\tau)u^+_\alpha$$
$$\rho^a(\sigma,\tau,u) = \rho^a(\sigma,\tau) + f_2^a(\sigma,\tau,u)$$
$$\eta^-_{-a}(\sigma,\tau,u) = \eta^\alpha_{-a}(\sigma,\tau)u^-_\alpha + g^-_{1-a}(\sigma,\tau,u) \quad (2.24)$$
$$\xi^+_{+a} = \xi^\alpha_{+a}(\sigma,\tau)u^+_\alpha + k^+_{1+a}(\sigma,\tau,u)$$
$$b^{--}_{--a}(\sigma,\tau,u) = b^{\alpha\beta}_{--a}(\sigma,\tau)u^-_\alpha u^-_\beta + L^{--}_{1--a}(\sigma,\tau,u)$$
$$C^{++}_{++a}(\sigma,\tau,u) = C^{\alpha\beta}_{++a}(\sigma,\tau)u^+_\alpha u^+_\beta + L^{++}_{2++a}(\sigma,\tau,u)$$

Thus, to first order in harmonic variables, the equations (2.21) become

$$\dot{\chi}^{-l}_+ + c\chi'^{-l}_+ + \frac{1}{2}\Gamma^l_{ak}(\omega)[\dot\omega^a \chi^{-k}_+ + C\omega'^a \chi^{-k}_+] = 0$$

$$\dot\psi^{+l}_- - c\psi'^{+l}_- + \Gamma^l_{ak}(\omega)[\dot\omega^k \psi^{+a}_- - c\psi^{+a}_- \omega'^k] = 0 \quad (2.25)$$

$$\ddot\omega^l - c^2\omega''^l + \Gamma^l_{ak}(\omega)[\dot\omega^l\dot\omega^k - c^2\omega'^l \omega'^k] = 0$$

and the equations obtained from the variation with respect to the two-dimensional metric $h^{--++}(z_A,u)$ take the following expressions

$$G_{ab}(\omega)[\partial_{--}\omega^a \chi^{-k}_+] = 0$$
$$G_{ab}(\omega)[\psi^{+a}_-\partial_{++}\omega^b] = 0$$
$$G_{ab}(\omega)[\psi^{+a}_-(\xi^{-b}_+ - 2\partial_{++}\psi^{-b}_-) + (\eta^{+a}_- - 2\partial_{--}\chi^{+a}_-)\chi^{-b}_+ - 2b^{+-a}_{--}\partial_{++}\omega^b \quad (2.26)$$
$$- 2(C^{+-b}_{++} - 2\partial_{++}\omega^b)\partial_{--}\omega^a] = 0$$

Let us recall that
$$\partial_{\pm\pm} = \partial_\tau \mu\, c\partial_\sigma$$
$$\frac{1}{2}(u^{+\alpha}u^{-\beta} + u^{+\beta}u^{-\alpha}) = \delta^{\alpha\beta}$$

Consequently, the above equations become

$$G_{ab}(\omega)[\dot\omega^a + c\omega'^a]\chi^{-b}_+ = 0 \quad (1)$$

$$G_{ab}(\omega)[\dot\omega^a - c\omega'^a]\psi^{+b}_- = 0 \quad (2) \quad (2.27)$$

$$G_{ab}(\omega)[\psi^{+a}_-(\xi^{-b}_+ - 2\dot\psi^{-b}_- + 2c\psi'^{-b}_-) + (\eta^{+a}_- - 2\dot\chi^{+a}_- - 2c\chi'^{+a}_-)\chi^{-b}_+$$
$$+ b^{+-a}_{--}(C^{+-b}_{++} - 2\dot\omega^b + 2c\omega'^b) - 2(C^{+-b}_{++} - 2\dot\omega^b + 2c\omega'^b)(\dot\omega^a + c\omega'^a)] = 0 \quad (3)$$

which imply the following constraints
$$G_{ab}(\omega)T^{ab} = 0$$
$$G_{ab}(\omega)[-\dot\omega^a\dot\omega^b + c^2\omega'^a \omega'^b] = 0 \quad (2.28)$$

with
$$T^{ab} = \omega'^a \dot\omega^b - c^2\dot\omega^a \omega'^b = -T^{ba}$$

and

$$G_{ab}(\omega) b_{--}^{\alpha\beta,a} \chi_+^{\gamma,b} = 0$$
$$G_{ab}(\omega) C_{++}^{\alpha\beta,b} \psi_-^{\gamma,a} = 0$$
$$G_{ab}(\omega)[(-\partial \omega^a + c\omega'^a)\chi_+^{\alpha,b}] = 0$$
$$G_{ab}(\omega)[(-\partial \omega^a + c\omega'^a)\psi_-^{\alpha,b}] = 0$$

(2.29)

The null superstring method consist of inserting into equations (2.25), (2.28) and (2.29) the expansion

$$\Omega^a(Z_A,u) = \Omega_1^a(Z_A,u) + c^2 \Omega_2^a(Z_A,u) + O(c^4) \quad (2.30)$$

with

$$\left|\Omega_2^a\right| \ll \left[\Omega_1^a\right]$$

which can be represented as

$$\omega^a(\tau,\sigma) = \omega_1^a(\tau,\sigma) + c^2 \omega_2^a(\tau,\sigma)$$
$$\psi_-^{\alpha,a}(\tau,\sigma) = \psi_{-1}^{\alpha,a}(\tau,\sigma) + c\psi_{-2}^{\alpha,a}(\tau,\sigma) \quad (2.31)$$
$$\chi_+^{\alpha,a}(\tau,\sigma) = \chi_{+1}^{\alpha,a}(\tau,\sigma) + c\chi_{+2}^{\alpha,a}(\tau,\sigma)$$

Therefore, at zeroth order in $c^2$ and in $c$ for $\omega^a$ and the spinorial fields respectively the equations (2.25) lead to

$$\dot{\chi}_{+1}^{\alpha,l} + \Gamma_{ak}^l(\omega)[\dot{\omega}_1^a \chi_{+1}^{\alpha,k}] = 0$$
$$\dot{\psi}_{-1}^{\alpha,l} + \Gamma_{ak}^l(\omega)[\dot{\omega}_1^a \psi_{-1}^{\alpha,k}] = 0 \quad (2.32)$$
$$\ddot{\omega}_1^l + \Gamma_{ak}^l(\omega)[\dot{\omega}_1^a \dot{\omega}_1^k] = 0$$

In the same way the equations (2.28) and (2.29) allow to obtain

$$\begin{aligned}
G_{ab}(\omega)\, \dot{\omega}_1^a\, \dot{\omega}_1^b &= 0 & (1)\\
G_{ab}(\omega)\, \omega'^a_1\, \dot{\omega}_1^b &= 0 & (2)\\
G_{ab}(\omega)\, \dot{\omega}_1^a\, \chi_{+1}^{\alpha,b} &= 0 & (3)\\
G_{ab}(\omega)\, \dot{\omega}_1^b\, \psi_{-1}^{\alpha,a} &= 0 & (4)\\
G_{ab}(\omega)\, \psi_{-1}^{\alpha,a}\, \dot{\psi}_{+1}^{\beta,b} &= 0 & (5)\\
G_{ab}(\omega)\, \dot{\chi}_{-1}^{\alpha,a}\, \chi_{+1}^{\beta,b} &= 0 & (6)\\
G_{ab}(\omega)\, \chi'^{\alpha,a}_{-1}\, \chi_+^{\beta,b} &= 0 & (7)\\
G_{ab}(\omega)\, \psi'^{\beta,b}_+\, \dot{\psi}_-^{\alpha,a} &= 0 & (8)
\end{aligned} \quad (2.33)$$

We observe that the coordinates $\omega_1^a(\tau,\sigma)$ and their supersymmetric partners $\psi_{-1}^{i,a}$ and $\chi_{+1}^{i,a}$ describe null superstring [7,12], a collection of points moving independently along null geodesics.
As in the ordinary case we find that the constraint (2.33.2) ensures that each point of the string propagates in a direction perpendicular to the string. The only reminiscence from superstring is the constraints (2.33.1- 4) which require the velocity to be perpendicular to the string and to their supersymmetric partners.

Furthermore, the next order corrections (first order in $c^2$ for $\omega_{2a}$ and first order in $c$ for $\psi_{-2}^{i,a}$ and $\chi_{+2}^{i,a}$) are given by

$$\omega_2^{'l} - \omega_1^{"l} + \Gamma_{ak}^l(\omega)[\omega_1^{'a}\omega_2^{'k} + \omega_2^{'a}\omega_1^{'k} - \omega_1^{'a}\omega_1^{'k}] + \frac{1}{2}\Gamma_{ak,m}^l(\omega)\omega_1^{'a}\omega_1^{'k}\omega_2^m = 0$$

$$G_{ab}(\omega)[\omega_1^{'a}\omega_2^{'b} + \omega_2^{'a}\omega_1^{'b} - \omega_1^{'a}\omega_1^{'b}] + G_{ab,l}(\omega)\omega_1^{'a}\omega_2^{'b}\omega_2^l = 0 \quad (2.34)$$

$$G_{ab}(\omega)[\omega_2^{'a}\omega_1^{'b} + \omega_1^{'a}\omega_2^{'b} - \omega_2^{'a}\omega_2^{'b}] + G_{ab,l}(\omega)[\omega_1^{'a}\omega_1^{'b} + \omega_1^{'a}\omega_1^{'b}]\omega_2^l = 0$$

$$\chi_{+2}^{'-l} + \Gamma_{ak,m}^l(\omega)\omega_1^{'a}\chi_{+1}^{-k}\omega_2^m + \Gamma_{ak}^l(\omega)(\omega_2^{'a}\chi_{+1}^{-k} + \omega_1^{'a}\chi_{+2}^{-k}) = 0$$

$$\psi_{-2}^{'+l} - \Gamma_{ak,m}^l(\omega)\omega_1^{'a}\psi_{-1}^{+a}\omega_2^m - \Gamma_{ak}^l(\omega)(\omega_2^{'k}\psi_{-}^{+a} - \psi_{-}^{+a}\omega_1^{'k}) = 0 \quad (2.35)$$

$$G_{ab,l}(\omega)\omega_2^l\omega_1^{'a}\chi_{+,1}^{i,b} + G_{ab}(\omega)[\omega_2^{'b}\chi_{+1}^{i,b} + \omega_1^{'a}\chi_{+2}^{i,b}] = 0$$

$$G_{ab,l}(\omega)\omega_2^l\omega_1^{'b}\psi_{+,1}^{i,a} + G_{ab}(\omega)[\omega_2^{'b}\psi_{-1}^{i,a} - \omega_1^{'b}\psi_{+2}^{i,a}] = 0 \quad (2.36)$$

$$\psi_{-,2}^{'i,l} - \psi_{-,1}^{'i,l} + \Gamma_{ak}^l(\omega)(\omega_1^{'k}\psi_{-2}^{i,a} - \psi_{-1}^{i,a}\omega_1^{'k}) = 0$$

$$\chi_{+,2}^{'i,l} + \chi_{+,1}^{'i,l} + \Gamma_{ak}^l(\omega)(\omega_2^{'k}\chi_{-2}^{i,a} - \chi_{-1}^{i,a}\omega_1^{'k}) = 0 \quad (2.37)$$

$$G_{ab}(\omega)[\omega_2^{'b}\chi_{+2}^{i,b} + \omega_1^{'a}\chi_{+1}^{i,b}] = 0$$

$$G_{ab}(\omega)[\omega_2^{'b}\psi_{-2}^{i,a} - \omega_1^{'b}\psi_{-1}^{i,a}] = 0 \quad (2.38)$$

where $\Gamma_{ab,l}^m$ and $G_{ab,l}$ indicate the derivatives with respect to $\omega^l$. In the following we will concentrate on just the constraints obtained at zeroth order in $c^2$ and in $c$ for $\omega^a$ and the spinorial fields respectively in order to quantize the null superstring in the de Sitter space.

### III - Quantization of (4,4) null superstring in de Sitter space

The line element in de Sitter space is defined as
$$dS^2 = c^2(\Omega_0)[d\Omega_0^2 - d\Omega_1^2 - d\Omega_2^2 - d\Omega_3^2] \quad (3.1)$$

with

$$c(\Omega_0) = \frac{R_0}{\Omega_0} \quad (3.2)$$

where $\Omega_0$ is the superfield corresponding to the superconformal time and $R_0$ is the scale factor. The (4,4) supersymmetric lagrangian for a null superstring propagating in (3.1) takes the following form

$$L = \int d^2\theta_+^+ d^2\theta_-^- du \, c^2(\Omega_0)[D^{++}\Omega_a D^{--}\Omega_b \eta^{ab}] \quad (3.3)$$

where

$$D^{++} = \partial^{++} - 2\theta_+^+\theta_+^+\partial_0$$
$$D^{--} = \partial^{++} - 2\theta_-^-\theta_-^-\partial_0$$

are the harmonic derivatives.

The lagrangian (3.3) can be expanded in terms of components at zero order in harmonic variable $u$ for the field $\omega$ and first order in $u$ for the spinorial fields $\psi$ and $\chi$ as follows

$$L = c^2(\omega_0)\left\{\dot{\omega}^2 - \frac{1}{2}\psi_{-}^{\alpha}\dot{\psi}_{+\alpha} - \frac{1}{2}\dot{\chi}_{-}^{\alpha}\chi_{+\alpha}\right\} \tag{3.4}$$

where

$$\dot{\omega}^2 = \dot{\omega}_0^2 - \sum_{i=1}^{3}\dot{\omega}_i^2$$

$$\psi_{-}^{\alpha}\dot{\psi}_{+\alpha} = \psi_{-0}^{\alpha}\dot{\psi}_{+0\alpha} - \sum_{j=1}^{3}\psi_{-i}^{\alpha}\dot{\psi}_{+\alpha i}$$

Furthermore, by using the constraint (2.33.2) the conjugate momentum of the fields $\omega^a$, with $a = 0,1,2,3$ are given by

$$P_0(\sigma) = c^2(\omega_0)\dot{\omega}_0 \tag{3.5.1}$$
$$P_i(\sigma) = c^2(\omega_0)\dot{\omega}_i \tag{3.5.2}$$

Similarly, the use of the constraints (2.33.5) and (2.33.6) leads to the conjugate momentums of the spinorial fields $\psi_{+a}^{\alpha}$ and $\chi_{-a}^{\alpha}$ which are respectively given by

$$Q_{-a}^{\alpha}(\sigma) = c^2(\omega_0)\psi_{-a}^{\alpha} \tag{3.6.1}$$
$$F_{+a}^{\alpha}(\sigma) = c^2(\omega_0)\chi_{+a}^{\alpha} \tag{3.6.2}$$

with $\alpha = 1, 2$ and we obtain the following constraints

$$\psi_{+,a}^{\alpha}(\sigma,\tau) = \overline{\psi}_{+,a}^{\alpha}(\sigma) \tag{3.7.1}$$
$$\chi_{-,a}^{\alpha}(\sigma,\tau) = \overline{\chi}_{-,a}^{\alpha}(\sigma) \tag{3.7.2}$$

On the other hand, we use the fact that

$$c(\omega_0) = \frac{R_0}{\omega_0} \tag{3.8}$$

Consequently, we deduce from the equation (3.5.1) that

$$\omega_0(\sigma,\tau) = \frac{R_0^2\,\overline{\omega_0}}{R_0^2 - P_0(\sigma)\overline{\omega_0}\,\tau} \tag{3.9}$$

where

$$\overline{\omega_0} = \omega_0(\sigma, \tau = 0)$$

In the same way, the equation (3.5.2) gives

$$\omega_i(\sigma,\tau) - \overline{\omega_i} = \frac{P_i(\sigma)\,\overline{\omega_0}^{\,2}\,\tau}{R_0^2 - \overline{\omega_0}\,P_0(\sigma)\,\tau} \tag{3.10}$$

with

$$\overline{\omega_i} = \omega_i(\sigma, \tau = 0), \qquad i = 1, 2, 3$$

Then, from the equations (3.6.1) and (3.6.2) we obtain respectively

$$\psi_{-a}^{\alpha}(\sigma,\tau) = \frac{R_0^2\,\overline{\omega_0}^{\,2}\,Q_{-a}^{\alpha}(\sigma)}{(R_0^2 - P_0(\sigma)\overline{\omega_0}\,\tau)^2} \tag{3.11}$$

$$\chi_{+a}^{\alpha}(\sigma,\tau) = \frac{R_0^2\,\overline{\omega_0}^{\,2}\,F_{+a}^{\alpha}(\sigma)}{(R_0^2 - P_0(\sigma)\overline{\omega_0}\,\tau)^2} \tag{3.12}$$

Therefore, the constraints (2.33.1) until (2.33.8) lead respectively to

$$P_0^2(\sigma) = \sum_{i=1}^{3} P_i^2(\sigma) \tag{3.13.1}$$

$$\overline{\omega}_0' P_0^2(\sigma) = \sum_{i=1}^{3} \overline{\omega}_i' P_i(\sigma) \tag{3.13.2}$$

$$P_0(\sigma) F_{+0}^\alpha(\sigma) = \sum_{i=1}^{3} P_i(\sigma) F_{+i}^\alpha(\sigma) \tag{3.13.3}$$

$$P_0(\sigma) Q_{-0}^\alpha(\sigma) = \sum_{i=1}^{3} P_i(\sigma) Q_{-i}^\alpha(\sigma) \tag{3.13.4}$$

$$\overline{\psi}_{+0}^\alpha(\sigma) Q_{-0}^\beta(\sigma) = \sum_{i=1}^{3} \overline{\psi}_{+i}^\alpha(\sigma) Q_{-i}^\alpha(\sigma) \tag{3.13.5}$$

$$\overline{\chi}'^{\alpha}_{=0}(\sigma) F_{+0}^\beta(\sigma) = \sum_{i=1}^{3} \overline{\chi}'^{\alpha}_{=i}(\sigma) F_{+i}^\beta(\sigma) \tag{3.13.6}$$

$$\overline{\chi}'^{\alpha}_{-0} \chi^{\beta}_{+0} = \sum_{i=1}^{3} \overline{\chi}'^{\alpha}_{=i} \chi^{\beta}_{+i} \tag{3.13.7}$$

$$\overline{\psi}'^{\alpha}_{+0} \psi^{\beta}_{-0} = \sum_{i=1}^{3} \overline{\psi}'^{\alpha}_{+i} \psi^{\beta}_{-i} \tag{3.13.8}$$

By using the expression (3.9) the expression (3.10) can be rewritten as

$$\omega_i - \overline{\omega}_i = \frac{P_i(\sigma)}{P_0(\sigma)}(\omega_0 - \overline{\omega}_0) \tag{3.14}$$

and for the supersymmetric partners we have

$$(\psi^\alpha_{-a} - \overline{\psi}^\alpha_{-a}) = \frac{Q^\alpha_{-a}(\sigma)}{R_0^2}(1+\frac{\omega_0}{\overline{\omega}_0})(\omega_0 - \overline{\omega}_0) \tag{3.15.1}$$

$$(\chi^\alpha_{+a} - \overline{\chi}^\alpha_{+a}) = \frac{F^\alpha_{+a}(\sigma)}{R_0^2}(1+\frac{\omega_0}{\overline{\omega}_0})(\omega_0 - \overline{\omega}_0) \tag{3.15.2}$$

As in the ordinary case [16] we note that previous relations are expressed in terms of the cosmic variables and we obtain straight lines for the individual particles of the null superstring. Furthermore, the expressions (3.14) and (3.15) kept their forms under the reparametrization $\tau = f(\tilde{\tau}, \tilde{\sigma})$ and $\sigma = g(\tilde{\sigma})$ where $f$ and $g$ are arbitrary functions. In fact, choosing [16]

$$\tau = \frac{R_0^2 \tilde{\tau}}{\overline{\omega}_0^2 + \overline{\omega}_0 P_0 \tilde{\tau}} \tag{3.16}$$

we find that

$$\omega_a = P_a \tilde{\tau} + \overline{\omega}_a \tag{3.17.1}$$

$$\psi^\alpha_{-a} = \frac{Q^\alpha_{-a}}{R_0^2}\left[\overline{\omega}_0 + 2P_0 \tilde{\tau} + \frac{P_0^2}{\overline{\omega}_0}\tilde{\tau}^2\right] \tag{3.17.2}$$

$$\chi^\alpha_{+a} = \frac{F^\alpha_{+a}}{R_0^2}\left[\overline{\omega}_0 + 2P_0 \tilde{\tau} + \frac{P_0^2}{\overline{\omega}_0}\tilde{\tau}^2\right] \tag{3.17.3}$$

On the other hand, the expansions of the variables $\bar{\omega}^a(\sigma)$, $\bar{\psi}^\alpha_{+a}(\sigma)$, $\bar{\chi}^\alpha_{-a}(\sigma)$ and their conjugate momentums in Fourier series are given as follows

$$\bar{\omega}^a(\sigma) = \sum_n x_n^a e^{in\sigma} \quad , \quad P^a(\sigma) = \sum_n p_n^a e^{in\sigma}$$

$$\bar{\psi}^{\alpha a}_+(\sigma) = \sum_n \bar{\zeta}^{\alpha a}_{+n} e^{in\sigma} \quad , \quad Q^{\alpha a}_-(\sigma) = \sum_n q^{\alpha a}_{-n} e^{in\sigma} \qquad (3.18)$$

$$\bar{\chi}^{\alpha a}_-(\sigma) = \sum_n \bar{\xi}^{\alpha a}_{-n} e^{in\sigma} \quad , \quad F^{\alpha a}_+(\sigma) = \sum_n f^{\alpha a}_{+n} e^{in\sigma}$$

Similarly, the constraints (3.13.1) until (3.13.6) can be expanded in Fourier series as

$$P^a P_a = \sum_n H_n e^{in\sigma} \quad , \quad \bar{\omega}'^a P_a = \sum_n G_n e^{in\sigma}$$

$$P^a Q^\alpha_{-a} = \sum_n Y^\alpha_{-n} e^{in\sigma} \quad , \quad \bar{\psi}^\alpha_{+a} Q^{\beta a}_- = \sum_n L^{\alpha\beta}_n e^{in\sigma} \qquad (3.19)$$

$$P^a F^\alpha_{+a} = \sum_n F^\alpha_{+n} e^{in\sigma} \quad , \quad \bar{\chi}^\alpha_{-a} F^{\beta a}_+(\sigma) = \sum_n M^{\alpha\beta}_n e^{in\sigma}$$

By substituting the expressions (3.18) in the above relations (3.19) we find that

$$H_k = \sum_n p_n p_{k-n} \quad , \quad F^\alpha_{+k} = \sum_n p_{k-n} f^\alpha_{+n}$$

$$G_k = i\sum_n n\, x_n\, p_{k-n} \quad , \quad L^{\alpha\beta}_k = \sum_n \bar{\zeta}^\alpha_{+n} q^\beta_{-(k-n)} \qquad (3.20)$$

$$Y^\alpha_{-k} = \sum_n p_{k-n} q^\alpha_{-n} \quad , \quad M^{\alpha\beta}_k = \sum_n \bar{\xi}^\alpha_{-n} f^\beta_{+(k-n)}$$

The string coordinates $\bar{\omega}_a(\sigma)$ and their supersymmetric partners $\bar{\psi}^\alpha_{+a}(\sigma)$ and $\bar{\chi}^\alpha_{-a}(\sigma)$ satisfy with the momentum conjugates $P^a(\sigma)$, $Q^\alpha_{-a}(\sigma)$ and $F^\alpha_{+a}(\sigma)$ respectively the following Poisson brackets

$$\begin{aligned}
\{P^a(\sigma)\,,\,\bar{\omega}_b(\sigma')\} &= \delta(\sigma-\sigma')\,\eta^a_b \\
\{Q^{\alpha a}_-(\sigma)\,,\,\bar{\psi}^\beta_{+b}(\sigma')\} &= \delta^{\alpha\beta}\,\delta(\sigma-\sigma')\eta^a_b \\
\{F^{\alpha a}_+(\sigma)\,,\,\bar{\chi}^\beta_{-b}(\sigma')\} &= \delta^{\alpha\beta}\,\delta(\sigma-\sigma')\,\eta^a_b
\end{aligned} \qquad (3.21)$$

while for their Fourier modes we find

$$\begin{aligned}
\{p_n^a\,,\,x_{mb}\} &= \delta_{n+m,0}\,\eta^a_b \\
\{q^{\alpha,a}_{-,n}\,,\,\bar{\zeta}^\beta_{+mb}\} &= \delta^{\alpha\beta}\,\delta_{n+m,0}\,\eta^a_b \\
\{f^{\alpha,a}_{+,m}\,,\,\bar{\xi}^\beta_{-mb}\} &= \delta^{\alpha\beta}\,\delta_{n+m,0}\,\eta^a_b
\end{aligned} \qquad (3.22)$$

Consequently, the Poisson brackets of the momentums (3.20) are given by

$$\begin{aligned}
\{G_n, G_m\} &= -i(n-m)\,G_{n+m} \\
\{H_n, H_m\} &= 0 \\
\{G_n, H_m\} &= i(n-m)\,H_{n+m} \\
\{L_n^{\alpha\beta}, L_m^{\alpha'\beta'}\} &= \delta^{\beta\alpha'}L_{n+m}^{\alpha\beta'} + \delta^{\alpha\beta'}L_{n+m}^{\alpha'\beta} \\
\{M_n^{\alpha\beta}, M_m^{\alpha'\beta'}\} &= \delta^{\beta\alpha'}M_{n+m}^{\alpha\beta'} + \delta^{\alpha'\beta}M_{n+m}^{\alpha\beta'} \\
\{Y_{-n}^{\alpha}, Y_{-m}^{\beta}\} &= 0 \\
\{F_{+n}^{\alpha}, F_{+m}^{\beta}\} &= 0 \\
\{F_{+n}^{\alpha}, Y_{-m}^{\beta}\} &= -\delta^{\alpha\beta}H_{n+m} \\
\{G_n, Y_{-m}^{\alpha}\} &= \left(\frac{n}{2}-m\right)Y_{-,n+m}^{\alpha} \\
\{G_n, F_{+m}^{\alpha}\} &= \left(\frac{n}{2}-m\right)F_{+,n+m}^{\alpha}
\end{aligned} \qquad (3.23)$$

Other relations can also be obtained. We note that $G_n$ and $H_n$ generate reparametrizations in $\sigma$ and $\tau$ respectively like in the ordinary case [16]. Furthermore, the (4,4) superalgebra obtained contains two independent SU(2) Kac-Moody algebras corresponding to the decomposition SO(4) ~ SU(2) × SU(2) [17].

Finally, we may now quantize the null (4,4) superstring by replacing the Poisson brackets by commutators. This leads to the (4,4) superalgebra which will takes the following form

$$\begin{aligned}
[G_n, G_m] &= (n-m)\,G_{n+m} + A(n)\,\delta_{n+m,0} \\
[H_n, H_m] &= B(n)\,\delta_{n+m,0} \\
[G_n, H_m] &= (n-m)\,H_{n+m} + C(n)\,\delta_{n+m,0} \\
\{Y_{-n}^{\alpha}, Y_{-m}^{\beta}\} &= \varphi_{--}(n)\,\delta_{n+m,0}\,\delta^{\alpha\beta} \\
\{F_{+n}^{\alpha}, F_{+m}^{\beta}\} &= \gamma_{++}(n)\,\delta_{n+m,0}\,\delta^{\alpha\beta} \\
\{F_{+n}^{\alpha}, Y_{-m}^{\beta}\} &= \delta^{\alpha\beta}H_{n+m} + a(n)\,\delta^{\alpha\beta} \\
[G_n, Y_{-m}^{\alpha}] &= \left(\frac{n}{2}-m\right)Y_{-,n+m}^{\alpha} \\
[G_n, F_{+m}^{\alpha}] &= \left(\frac{n}{2}-m\right)F_{+,n+m}^{\alpha} \\
[L_n^{\alpha\beta}, L_m^{\alpha'\beta'}] &= \delta^{\beta\alpha'}L_{n+m}^{\alpha\beta'} + \delta^{\alpha\beta'}L_{n+m}^{\alpha'\beta} + b(n)(\delta^{\alpha\alpha'}\delta^{\beta\beta'} - \delta^{\alpha\beta'}\delta^{\beta\alpha'})\delta_{n+m,0} \\
[M_n^{\alpha\beta}, M_m^{\alpha'\beta'}] &= \delta^{\alpha'\beta}M_{n+m}^{\alpha\beta'} + \delta^{\alpha\beta'}M_{n+m}^{\alpha'\beta} + d(n)(\delta^{\alpha\alpha'}\delta^{\beta\beta'} - \delta^{\alpha\beta'}\delta^{\beta\alpha'})\delta_{n+m,0}
\end{aligned} \qquad (3.24)$$

In order to specify the form of the anomalies, the Jacobi identities for the (4,4) superalgebra has to be used. Such calculation depends crucially upon the ordering of the operators [18]. The determination of the most general form of the (4,4) supersymmetric anomalies is under investigation.

## IV- Conclusion

The aim of these lectures is to extend the investigation of the bosonic string to the superstring in curved spacetime. We think that it will be helpful and necessary for the understanding of quantum supergravity. However, we have considered a null (4,4) superstring evolution and propagation in curved harmonic superspace. Therefore, the equation of motion and the constraints were solved by means of the null (4,4) superstring expansion. In this scheme, the superstring equations of motion and the constraints are systematically expanded in powers of the super world-sheet speed of the light $c^2$ which is proportional to the string tension $T_0$. The points of the null superstring interact only with the gravitational background and the analytical expressions of the zeroth and first order solutions are derived. Furthermore, we have quantized the (4,4) null superstring in de Sitter space and formulating the corresponding (4,4) super algebra.

## Acknowledgments

The authors would like to thank the International Atomic Energy Agency and UNESCO hospitality at the Abdus Salam International Centre for Theoretical Physics, Trieste. This work is supported by the framework of the Associate and Federation Schemes of the Abdus Salam International Centre for Theoretical Physics, Trieste, Italy.